\documentclass[12pt]{article}
\usepackage{pdproc,epsfig} 

\begin{document}

\renewcommand{\thefootnote}{\alph{footnote}}
  
\title{
WHAT IS THE ROLE OF NEUTRINOS IN SHAPING THE UNIVERSE?}
\author{ LAWRENCE M. KRAUSS}
\address{Departments of Physics and Astronomy,  
Case Western Reserve University \\
10900 Euclid Ave. 
Cleveland, OH~~44106-7079, USA\\
 {\rm E-mail: krauss@cwru.edu}}

\abstract{ I review various aspects of the role neutrinos have played in shaping various cosmological observables:  the nature of large scale structure, observed fluctuations in the CMB, the nature of matter, and the shape of things to come.   (Invited review lecture III International Workshop on NO-VE, Venice, 2006)}
   
\normalsize\baselineskip=15pt

\section{Introduction:}

I was asked by the organizers to speak on the role of neutrinos in shaping the universe.  Since I had no idea what this phrase meant, I was able to liberally interpret it.   I confess that for a moment I worried  that it might somehow refer to some exotic compactification of a higher dimensional space that might, in some Landscape model, be related to some fundamental flux associated with neutrinos.  But, checking that the rest of the meeting was firmly grounded in four dimensional physics, I happily decided that it must relate to something else.   

To aid me in my considerations, I turned to a previous review talk I gave in Venice a decade ago, in which I displayed images of the Universe near t=0, and today, as shown below:

\begin{figure}[htb]
\vglue 2.8in
\includegraphics{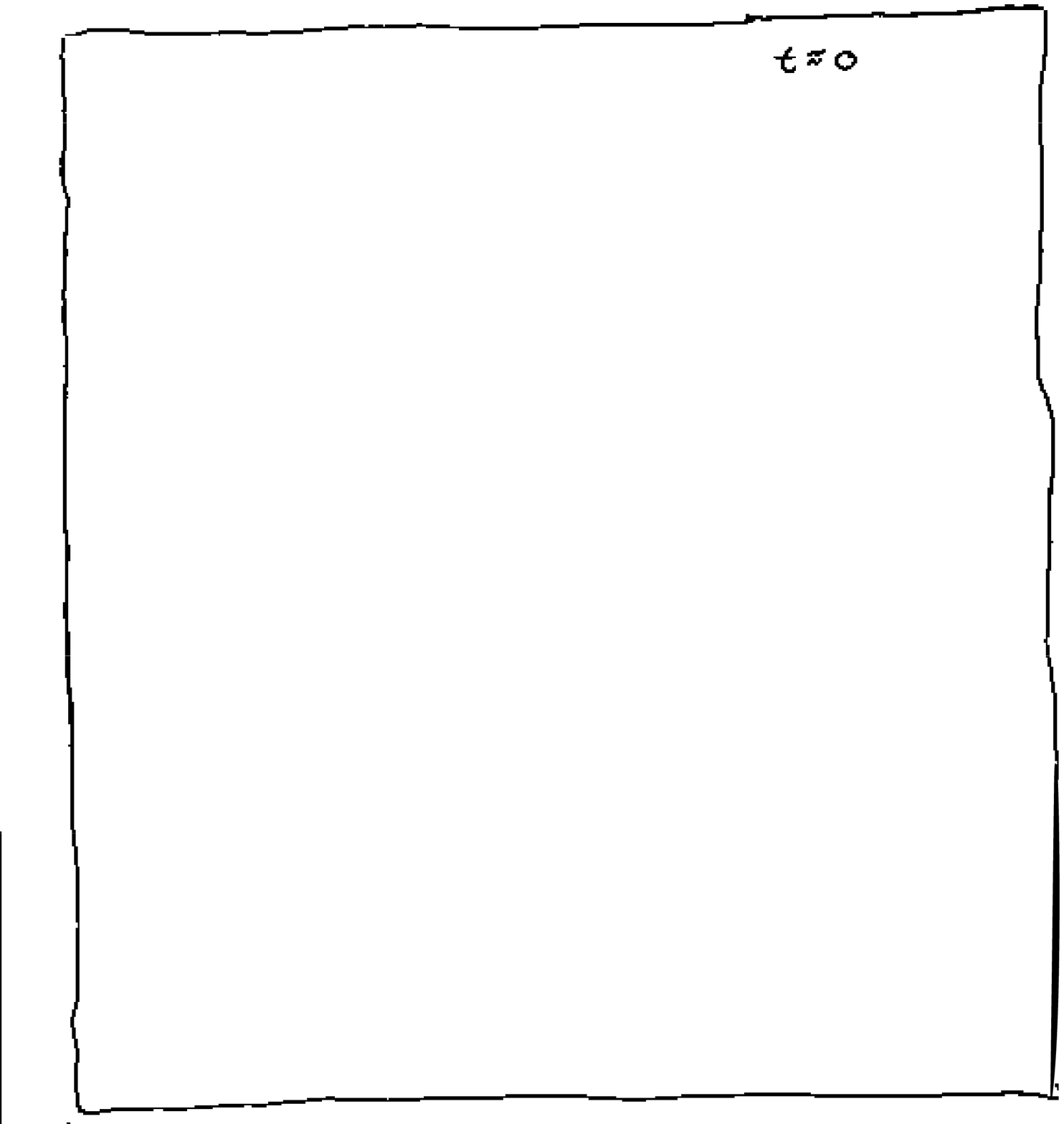}
\includegraphics{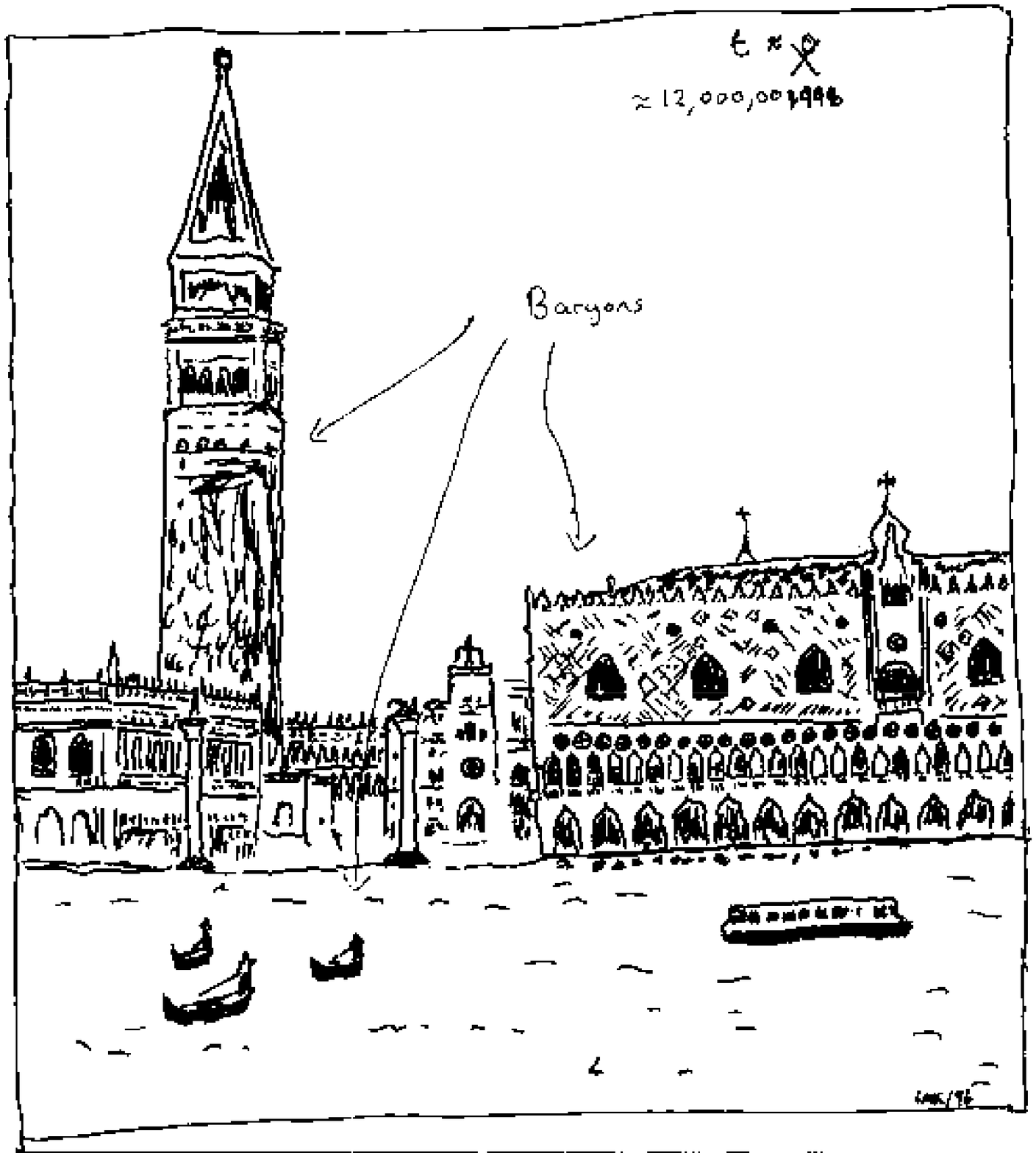}
\caption{(a) The Universe near t=0, \ \ \ \ \ \ \ \ \  (b) The Universe near today}
\label{fig:cmb2}
\end{figure}

Clearly the shapes in the second image are more interesting.   Thus, I took the charge of this talk to be an overview of how neutrinos have played a role in shaping various observables within Universe we see today, which is what I will attempt to summarize below.

\section{Neutrinos and the Shape of Structure:}

One of the most beautiful grand syntheses in physics in the late 1970's and early 80's was the realization that (a) galaxies appeared to be dominated by mysterious dark matter, and (b) if neutrinos had a small mass, then they could naturally be the dark matter, and (c) neutrinos appeared to have just the small mass needed!   The only problem with this grand synthesis was that it was wrong.   

The first inkling of this came as cosmologists began to examine Universea in which an initial flat adiabatic spectrum of gaussian density perturbations evolved into nonlinear structures on a computer.  N-body codes at the time allowed them to put in hot neutrino dark matter, as would be expected from 30 eV neutrinos, and the results\cite{davis} did not resemble at all recent redshift surveys of the galaxy distribution.

\begin{figure}[htb]
\vglue 3.2in
\includegraphics{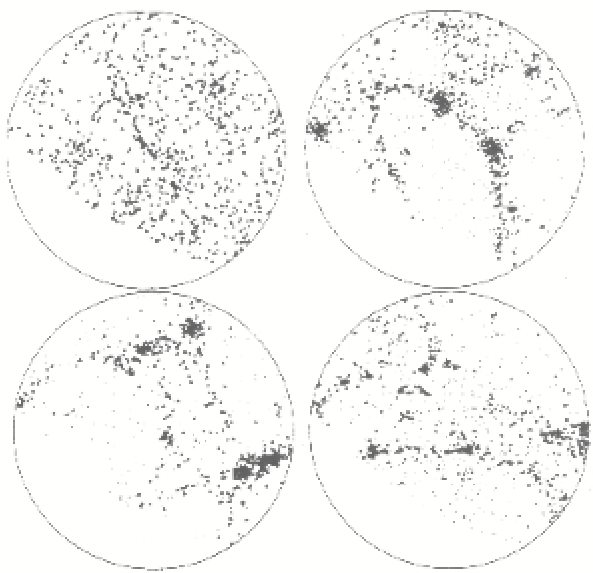}
\caption{ A slice of the nearby Universe as it looked in 1983 (upper left) vs the galaxy distribution in universes dominated by light neutrinos created on a computer  (from Davis et al (1985)}
\label{fig:neutr1}
\end{figure}

At around the same time, it became clear that the early experimental evidence from tritium beta decay for a non-zero neutrino mass had been wrong, and subsequent double beta decay experiments demonstrated that the neutrino masses were much smaller than 30 eV.  This is reflected in a recent Danish cartoon about the poor burned experimentalists: 

\begin{figure}[htb]
\vglue 2.8 in
\includegraphics{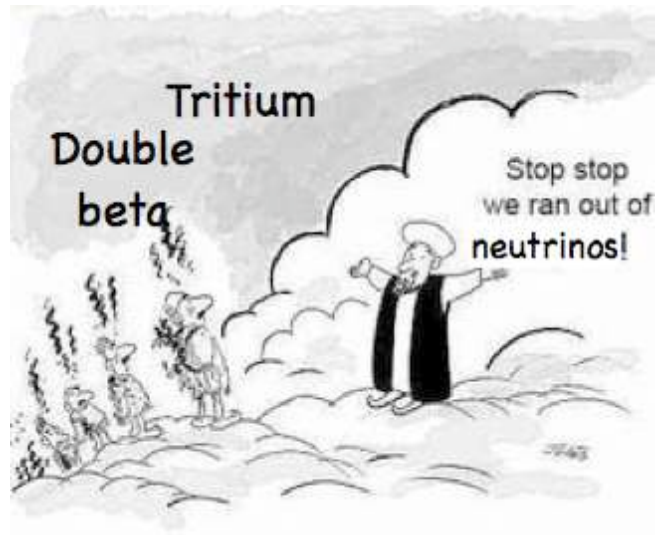}
\caption{Burned experimenters finding out that there are no light neutrinos left as dark matter}
\label{fig:neutr2}
\end{figure}

This sad tale was not for nought, however, because it provided a basic understanding of those characteristics that would be required for dark matter to produce observed structure, and led to the birth of Cold Dark Matter.  Let me review them briefly here.  Most of these arguments are completely independent of Inflation, although Inflationary models certainly provide first principles mechanisms for initiating the seeds of large scale structure.

We start by considering the primordial power spectrum of density perturbations.  Presented in momentum space, it is first natural to assume that there is no preferred scale, so that the power spectrum is {\it scale independent} and can be presented as a power law:

\begin{equation}
\left|{ \left( { \delta \rho \over \rho}\right)}_k \right|^2  \propto k^n
\end{equation}

Next, we know that the exponent $n \approx 1$.   While this is a consequence of most inflationary models, it is also an a priori requirement, because if $ n >> 1$ then the spectrum will blow up for large $k$ (i.e. small scales), and there were be too many primordial black holes in the Universe.   If $ n<<1$ then there is too much power for small $k$ (large scales), and the observed isotropy on large scales would be violated.

So much for the primordial spectrum.  Observed density fluctuations involve processing the primordial spectrum via gravitational instability, which in turn depends upon the equation of state of the dominant contribution to the energy density of the universe at any time.   In a radiation dominated universe density fluctuations do not grow inside the horizon.  In general they tend to damp out.   Since large $k$ corresponds to smaller scales, which come inside the horizon earlier, when the universe is radiation dominated, these tend to damp out.   Thus, the primordial spectrum gets processed to look like :

\begin{figure}[htb]
\vglue 2.4 in
\includegraphics{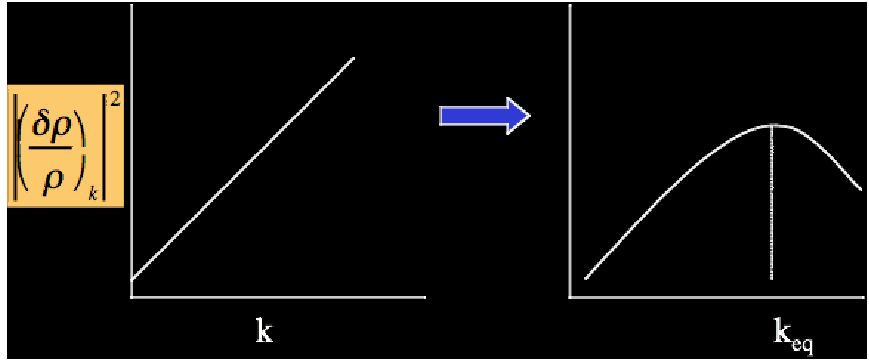}
\caption{Processing of the primordial power spectrum as fluctuations come within the horizon}
\label{fig:powers}
\end{figure}

In this figure $k_{eq}$ represents the wavenumber associated with a scale that comes inside the horizon at the time matter begins to dominate the universe.  For all smaller scales, fluctuations have damped as they came inside the horizon during radiation domination.   Thus $k_{eq}$  is proportional to the redshift of matter-radiation equality, which is itself determined by $\Omega_{matter} h^2$ today.   Now, since wavenumber is related to physical scale $ R$ by $ k^{-1} \propto Rh$, then the physical scale associated with the turnover is given by $R_{turnaround} \propto \Omega_{matter} h$.

Over the past decade, we have been fortunate to measure the density power spectrum on scales ranging from galaxies and smaller, via the Lyman alpha forest, to scales approaching the horizon size today, via the CMB.   A recent compilation from Tegmark\cite{Tegmark}  is shown below:

\begin{figure}[htb]
\vglue 2.6 in
\includegraphics{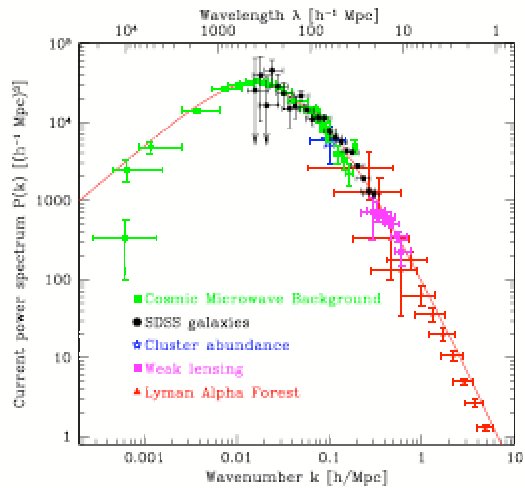}
\caption{Measurements of the density power spectrum today over a wide variety of scales}
\label{fig:powerteg}
\end{figure}

Now we can understand the effect of light neutrinos.   If neutrinos are lighter than an eV or so, they will be relativistic at the time when galaxy-sized fluctuations first come within the horizon, and they will have the effect of pushing the era of radiation-matter equality to later times.  This will have the effect of damping fluctuations on larger scales than would otherwise be damped.   There are two ways that neutrinos might thus damp power.  First, if the number of standard model light neutrinos were greater, then their net contribution to radiation at early times would be larger.  Alternatively, if the neutrino mass were greater then their contribution to the energy density today would be larger.   Working back, this implies that their contribution to the radiation density at early times would be larger.  These effects are shown below\cite{bell,hu}:

\begin{figure}[htb]
\vglue 2.4 in
\includegraphics{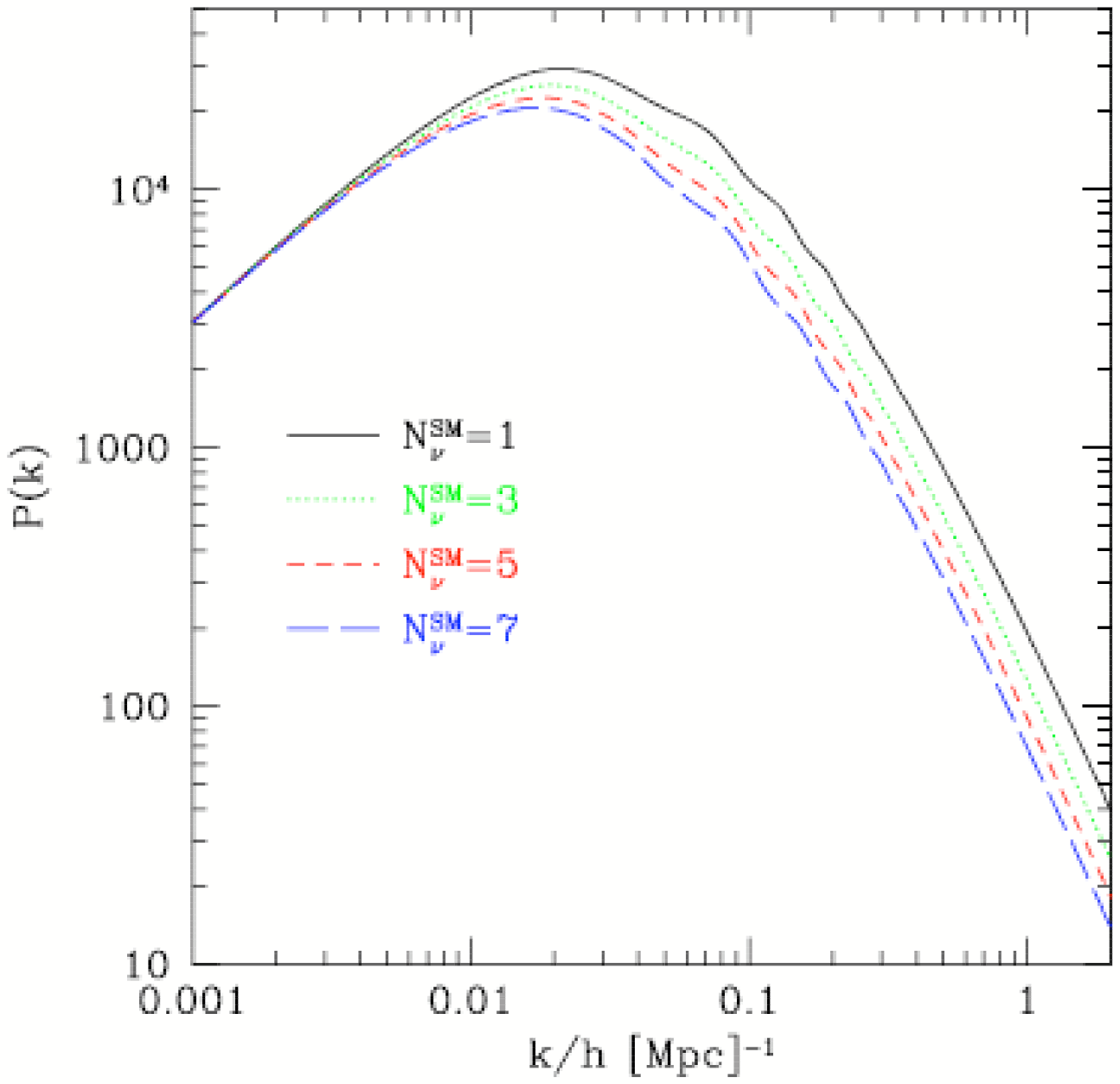}
\includegraphics{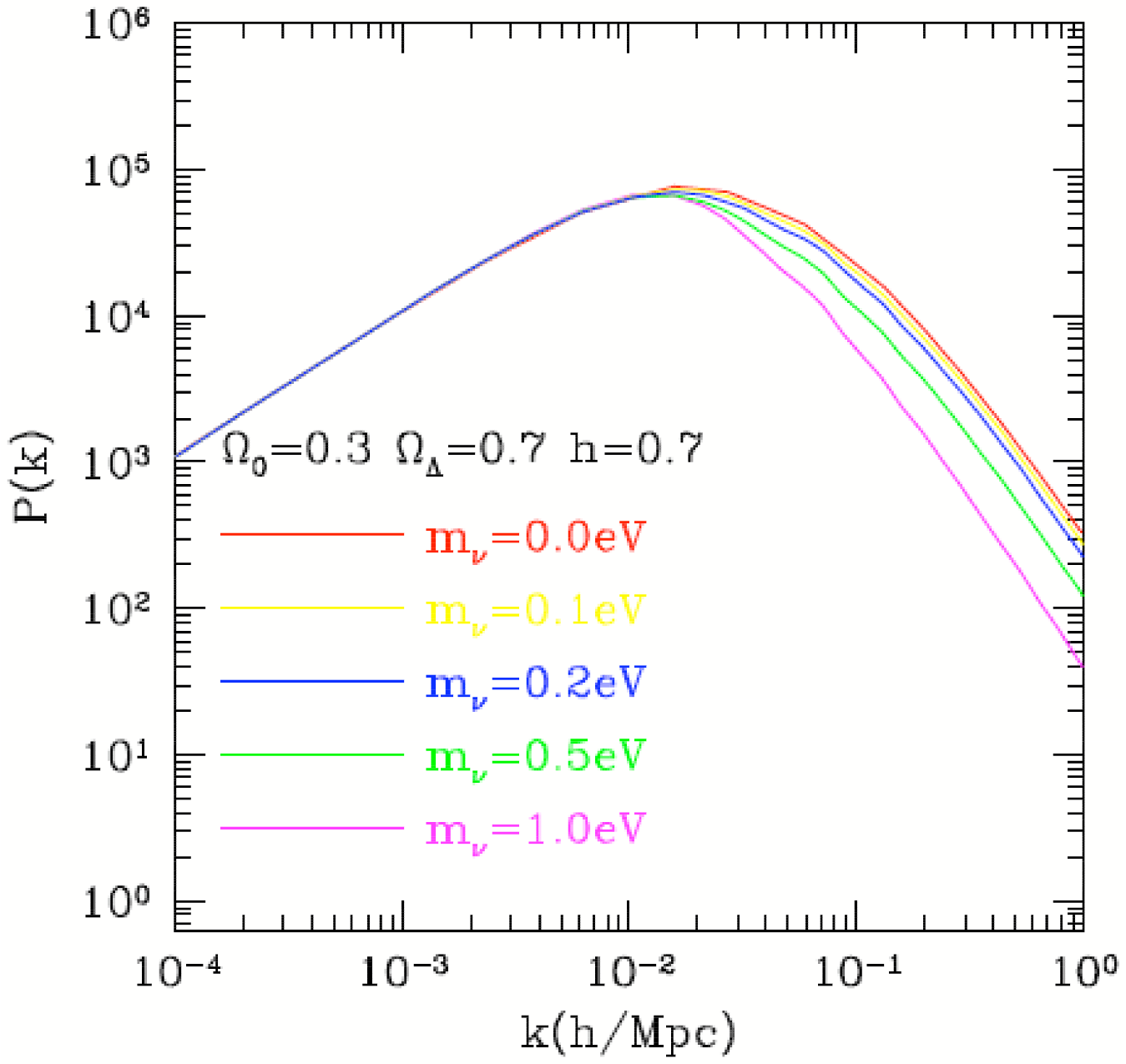}
\caption{Effects of light or massless neutrinos on density power spectrum}
\label{fig:neutrinopower}
\end{figure}

By comparing these effects with the observed power spectrum in Fig. 5, a limit on neutrino masses in the range of $0.17-0.6 $ eV has been established (i.e. see\cite{seljak}.

Because the Cosmic Microwave Background probes the nature of density fluctuations on the surface of last scattering at $z=1000$, neutrinos can affect the shape of the power spectrum of temperature fluctuations observed in the CMB.  Several different effects can occur.  

The first peak in the CMB power spectrum, the so-called acoustic peak arises from photons that have climbed out of large potential wells.  In the center of the wells, $\delta T/T$ is large and positive.  The effect of climbing out of the wells results in the so-called Sachs-Wolfe effect, which, due to the induced redshift on the photons moderates the $\delta T/T$ peak somewhat.

The effect of adding additional massless neutrino species, or giving neutrinos a small mass is to increase the energy density of relativistic particles at the time of last scattering.  The effect of these particles is to reduce the growth of fluctuations at that time (both by equation of state effects, and free-streaming out of potential wells).  The effect is to reduce the depth of the potential wells somewhat, and thus to reduce the impact of the Sachs-Wolfe redshifting, producing a bigger observed $\delta T/T$ associated with the first peak.   

At the same time, because the ratio of radiation to matter has changed in the universe, the time-temperature relationship will be changed, so that the epoch at which $z=1000$ will correspond to a different time, and horizon sized regions will correspond to different angular sizes on the microwave background.  This has the effect of shifting the peaks, especially the higher order peaks.  Both of these effects can be seen in the figure below, and can be used to constrain the energy density in light neutrinos from the CMB, thus constraining both neutrino masses, and the number of light families\cite{bell}.

\begin{figure}[htb]
\vglue 2.6 in
\includegraphics{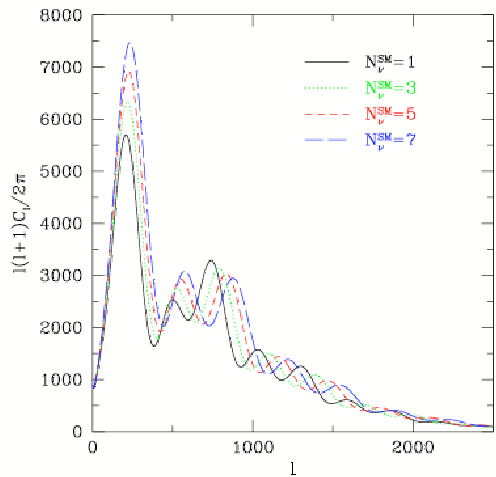}
\caption{Effect on CMB power spectrumof adding additional massless neutrino species }
\label{fig:neutcmb}
\end{figure}

\section{Neutrinos and the Shape of Matter}

In many and varied ways neutrinos are responsible for the existence of the matter we see around us today.   This is because (1) they dramatically impact upon the nuclear physics processes that govern the formation of light elements, and (2) also stellar dynamics and the infusion of heavy elements into the interstellar medium.  In addition, (3)  they may be responsible for the baryon asymmetry of the universe, and (4) even for cold dark matter!

\subsection{Big Bang Nucleosynthesis}

Weak interactions which mediate neutron-proton transitions play a vital role in the process of Big Bang Nucleosynthesis, by preserving thermal number densities of neutrons relative to protons, which of course involves the production and absorption of neutrions.  Were it not for the freeze-out of these interactions, no primordial neutrons would be left to seed the formation of helium during this epoch.  Neutrinos play a well known additional role in BBN, by contributing to the energy density of radiation that governs the expansion rate at that time.  More neutrino species means faster expansion, which means earlier freeze-out, which means more neutrons, and thus more helium.   Fewer neutrinos would mean less helium.   If there were much less helium in stars than the 25 $\%$ or so that was produced during BBN, the evolution of the first generation of stars would have been quite different.  I expect that we would not have expected to find as much heavy element abundance as we observe in our solar system, and with it the building blocks of life, and of course much of the rest of the matter we see around us, no matter what shape.

\subsection{Stellar Evolution and Supernovae}

Neutrinos actually play a far more important role in resulting in the heavy element abundance that we observe today.  Were it not for neutrino interactions, in fact, it is quite possible that no heavy elements might escape from early generations of stars to see our own solar system.  This is because it appears that  supernova explosions occur in part because the outgoing neutrino wind, which carries most of the energy emitted by the collapsing star, interacts with the dense material in a stalled outgoing shock layer, imparting enough energy to re-energize it and complete the explosion.   Were it not for this injection of energy, the processed heavy elements inside the star would not be injected into the interstellar medium, to one day be amalgamated into material that will collapse into other stars and solar systems.   Since essentially all elements heavier than lithium were produced in stellar furnaces and not in the Big Bang, without this injection, essentially everything we see on earth would not exist.   Interestingly, it is precisely because of the neutral current interactions of neutrinos that this is possible, because for MeV energy neutrinos these interactions can be coherent across an entire nucleus, giving a cross section that grows roughly as $A^2$, where $A$ is atomic number.  As a result, the cross sections can be much higher.   While this possibility has been known and discussed for over 30 years, coherent neutral current interactions of neutrinos, so vital to our existence, have not yet been directly measured in the laboratory.  Several ongoing experiments are currently underway to do this, as discussed at this meeting.

\subsection{Invisible Shapes:  A Neutrino Filled Universe}

Interestingly, because supernovae emit much more energy in neutrinos than they do in light, it turns out that the net effect of all supernovae that have occurred since the Big Bang (about 200 million in our galaxy alone!) is to have produce an additional cosmological neutrino background.  As we first estimated over 20 years ago \cite{krauss}, about 10-50 antineutrinos per square centimeter per second are going through each of us at this moment due to this background.   Recently, the possibility of detecting this background has again arisen as we have recently detected another antineutrino background: that due to radioactive decays within the Earth.  Just for fun, I display here a "grand unified" neutrino spectrum, containing all known and predicted neutrino backgrounds in the universe.  Once all of these were invisible, but perhaps by the end of this century, all will have been detected, giving us exciting new windows on otherwise invisible astrophysical sources in the universe.

\begin{figure}[htb]
\vglue 2.6 in
\includegraphics{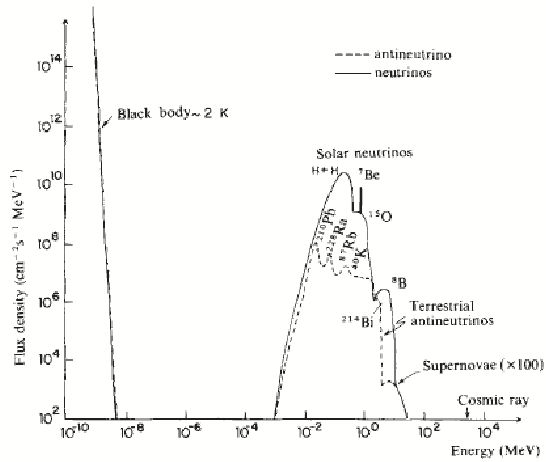}
\caption{The Grand Unified Cosmic Neutrino Spectrum on Earth }
\label{fig:neutgrandu}
\end{figure}

\subsection{Leptogenesis}

Ever since Andrei Sakharov described the conditions that were required in order for a baryon asymmetry to result from an initially particle-antiparticle symmetric universe, particle physicists have explored models that might naturally allow the three conditions he prescribed: (1) departure from thermal equilibrium, (2) CP violation, and (3) baryon violating processes.  This effort was made more difficult when it was recognized that non-perturbative weak interaction processes involving sphaelerons could, at a finite temperature, wash out any previously generated baryon asymmetry.  In recent years, however this problem has been turned into an advantage as it has been realized that the neutrino sector itself may contain all of the requirements to ultimately generate a baryon asymmetry.   If there are heavy Majorana neutrino masses, these violate lepton number.  If the neutrinos  decay out of equilibrium, and if there is CP violation in the neutrino mass matrix, which there can certainly be, then these decays can generate a lepton asymmetry in neutrinos.  But then since sphaelerons violate lepton number and baryon number separately, but not $B-L$, any initially generated lepton asymmetry could ultimately get turned into a baryon asymmetry.  In this way, neutrinos would be {\it truly} responsible for the shape of all matter we see in the universe today!

\subsection{Dark Matter}

While I described in the beginning of this article how it is that light neutrinos are ruled out as dark matter, the question naturally arises whether it might be possible that additional, heavy neutrino states might be viable cold dark matter candidates.  It has been known since the late 1980's after Z width measurements constrained the number of neutrinos with mass less than one half of the Z mass, that new standard model neutrinos, whose remnant abundance could be calculated by standard annihilation mechanisms in the early universe, are ruled out \cite{krauss89}.   However, what about the possible heavy Majorana neutrinos that are presumably a part of nature if a see-saw mechanism is used to generate the observed light neutrino masses?   In general these masses are quite heavy in order for the see-saw mechanism to give left-handed neutrinos light masses.   However, just to show that this is not absolutely essential, my collaborators and I showed some time back \cite{krausstrodden} that one can invent a very ugly model in which neutrinos only get masses from higher order loop effects, and in this case the Majorana states could be weak-scale, and still remain viable, and in fact also have abundances which are appropriate to be dark matter today.   Thus, there remains hope, however slim, that neutrinos are not only responsible for the generation of the shape of all observed matter, but also that of dark matter as well.  However, unfortunately in these models the right-handed neutrinos will be virtually undetectable in WIMP detection experiments.  Since I prefer experimental confirmation rather than theoretical uncertainty, I am not sure I am rooting for this possibility, even if I may have helped propose it. 

\section{Neutrinos and the Shape of Things to Come}

The single most significant development in cosmology in recent years has been the discovery of dark energy which is dominating the expansion of the universe.  Not only is it completely mysterious, but we now recognize that the nature of the dark energy will determine the future expansion history of the universe, independent of geometry and mass density today.  

In this regard, it has not been lost on many people that there is a remarkable coincidence in nature. exemplified by the following relation:

\begin{equation}
\rho_{\Lambda} = \Lambda^{1/4} \approx m_{\nu} !
\end{equation}

The energy density in dark energy today is absurdly small by comparison to all fundamental parameters in particle physics, {\it except} light neutrino masses!   Does this suggest perhaps that there might be some relationship between the observed energy of empty space and the physics of neutrinos?

One such suggestion, which has also been around for a long time, involves the possibility that what we are observing today as dark energy is really latent heat associated with a first  order phase transition that has not yet been completed.  If the latent heat is associated with a potential, schematically shown below, associated with a scalar field which undergoes spontaneous symmetry breaking, giving neutrinos mass in the process, then one might have a natural explanation of the coincidence in scales given above. 

\begin{figure}[htb]
\vglue 1.2 in
\includegraphics{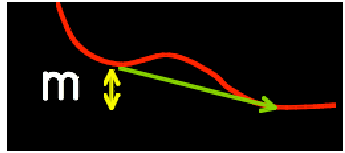}
\caption{A schematic potential involved in neutrino mass and dark energy generation }
\label{fig:neutdark}
\end{figure}

That is the good news.  The bad news is that if we are stuck in a metastable state with latent heat, then we are doomed eventually to experience a phase transition in the ground state of nature.   Such a phase transition could have dramatic and tragic implications, if not for the shape of the universe, for the shape we find ourselves in afterwards.   In this case, we will have a limited amount of time left to figure out the answer to the puzzles of neutrino masses, and dark energy.  Which means that we need to continue to have these delightful meetings in Venice as often as possible!

\section{Acknowledgements}
I thank Milla Baldo Ceolin for organizing once again such a delightful meeting, and also all neutrino experimentalists who continue to work against the odds in order to shed empirical light on the mysteries of particle physics in a way that can actually guide theorists.

\end{document}